\documentstyle[11pt,epsfig]{article}
\date{}
\begin{document}
\title{{\bf Quantization of the Schwarzschild black hole: a Noether symmetry approach}}
\author{Babak Vakili\thanks{email: b-vakili@iauc.ac.ir}\,\,\,\\\\{\small {\it Department of physics, Azad University of Chalous,  P.O. Box 46615-397, Chalous,
Iran,}}} \maketitle

\begin{abstract}
We study the canonical formalism of a spherically symmetric
space-time. In the context of the $3+1$ decomposition with respect
to the radial coordinate $r$, we set up an effective Lagrangian in
which a couple of metric functions play the role of independent
variables. We show that the resulting $r$-Hamiltonian yields the
correct classical solutions which can be identified with the
space-time of a Schwarzschild black hole. The Noether symmetry of
the model is then investigated by utilizing the behavior of the
corresponding Lagrangian under the infinitesimal generators of the
desired symmetry. According to the Noether symmetry approach, we
also quantize the model and show that the existence of a Noether
symmetry yields a general solution to the Wheeler-DeWitt equation
which exhibits a good correlation with the classical regime. We
use the resulting wave function in order to (qualitatively)
investigate the possibility of the
avoidance of classical singularities.\vspace{5mm}\noindent \vspace{0.8mm}\newline
\end{abstract}
\section{Introduction}
Black hole physics has played a central role in conceptual
discussion of general relativity at the classical and quantum
levels. For example regarding event horizons, space-time
singularities and also studying the aspects of quantum field
theory in curved space-time. In classical point of view, the
horizon of a black hole which is a one way membrane, and the
space-time singularities are some interesting features of the
black hole solutions in general relativity \cite{1}. In spirit of
the Ehrenfest principle, any classical adiabatic invariant
corresponds to a quantum entity with discrete spectrum, Bekenstein
conjectured that the horizon area of a non extremal quantum black
hole should have a discrete eigenvalue spectrum \cite{Bek}. Also,
the black hole thermodynamics is based on applying quantum field
theory to the curved space-time of a black hole \cite{Haw}.
According to this formalism, the Hawking radiation of a black hole
is due to random processes in the quantum fields near the horizon.
The mechanism of this thermal radiation can be explained in terms
of pair creation in the gravitational potential well of the black
hole \cite{Brl}. The conclusions of the above works are that the
temperature of a black hole is proportional to the surface gravity
and that the area of its event horizon plays the role of its
entropy. In this scenario, the black hole is akin to a
thermodynamical system obeying the usual thermodynamic laws, often
called the laws of black hole mechanics, first formulated by
Hawking \cite{Haw}. In more recent times, this issue has been at
the center of concerted efforts to describe and make clear various
aspects of the problem that still remain unclear, for a review see
\cite{Pad}. With the birth of string theory \cite{Seib}, as a
candidate for quantum gravity and loop quantum gravity
\cite{Asht}, a new window was opened to the problem of black hole
radiation. This was because the nature of black hole radiation is
such that quantum gravity effects cannot be neglected \cite{Muk}.
According to all of the above remarkable works, it is believed
that a black hole is a quantum mechanical system and thus like any
other quantum system its physical states can be described by a
wave function. Indeed, due to its fundamental conceptual role in
quantum general relativity, we may use it as a starting point for
testing different constructions of quantum gravity \cite{2,Kief}.

In this paper we deal with the Hamiltonian formalism of a static
spherically symmetric space-time. We show that the classical
solution of such a system can be identified with the space-time of
a Schwarzschild black hole. In this model a couple of metric
functions play the role of independent variables which with their
conjugate momenta construct the corresponding phase space. We then
study the existence of Noether symmetry in this phase space by
utilizing the behavior of the corresponding Lagrangian under the
infinitesimal generators of the desired symmetry. By the Noether
symmetry of a given phase space we mean that there exists a vector
field $X$, as the infinitesimal generator of the symmetry on the
tangent space of the configuration space such that the Lie
derivative of the Lagrangian with respect to this vector field
vanishes. Since the existence of a symmetry results in a constant
of motion (Noether charge), we show that the mass of the black
hole plays the role of the Noether charge in black hole system.
Finally, we consider a canonical quantum theory by replacing the
classical phase space variables by their Hermitian operators. We
study the various aspects of the resulting quantum model and
corresponding Wheeler-DeWitt equation and present closed form
expressions for the wave function of the black hole.
\section{The model}
We start with the Einstein-Hilbert action
\begin{eqnarray}\label{A}
{\cal S}=\frac{1}{16\pi G}\int d^4x \sqrt{-g}{\cal R}+S_{YGH},
\end{eqnarray}
where ${\cal R}$ is the Ricci scalar, $g$ is the determinant of
the metric tensor and $S_{YGH}$ is the York-Gibbons- Hawking
boundary term. Following \cite{3} we assume that the geometry of
space-time is described by a general static spherically symmetric
line element as \cite{4}
\begin{equation}\label{B}
ds^2=-a(r)dt^2+N(r)dr^2+2B(r)dtdr+b^2(r)(d\vartheta^2+\sin^2
\vartheta d\varphi^2),\end{equation} where $a(r)$, $B(r)$, $N(r)$
and $b(r)$ are undetermined functions of $r$. If we introduce a
new radial coordinate by the transformation $b(r)\rightarrow r'$
and define a new time coordinate by
$I(r)[a(r)dt-B(r)dr]\rightarrow dt'$, it easy to show that the
line element (\ref{B}) (after dropping the primes) takes the
standard form
\begin{equation}\label{C}
ds^2=-A(r)dt^2+C(r)dr^2+r^2(d\vartheta^2+\sin^2 \vartheta
d\varphi^2),\end{equation}in which there are two unknown functions
$A(r)$ and $C(r)$ to be determined by the Einstein field
equations. Although, the line element (\ref{C}) is the standard
spherically symmetric metric line element in four dimension, as in
\cite{3} our starting point to construct the Hamiltonian formalism
of the model will be the line element (\ref{B}). Indeed, we shall
see that a combination of the metric functions in (\ref{B}) forms
a Lagrange multiplier as
\begin{equation}\label{D}
n(r)=a(r)N(r)+B^2(r).\end{equation} On the other hand, in the
metric (\ref{B}), $N(r)$ plays the role of a lapse function with
respect to the $r$-slicing in the ADM terminology. Therefore,
since the functions $N(r)$ and $B(r)$ are related to the Lagrange
multiplier $n(r)$ by (\ref{D}), they can be arbitrarily chosen.
These two functions represent the freedom in the definition of the
coordinates $r$ and $t$ in the metric (\ref{B}). Hence, we are
left with two functions $a(r)$ and $b(r)$ which can be determined
by the Einstein field equations.

Now, let us deal with the canonical structure of the model at
hand. Putting equation (\ref{B}) in the expression (\ref{A}), we
find that the action for the gravitational field can be written as
\cite{3}
\begin{equation}\label{E}
{\cal S}=\int dt \int dr {\cal L}(a,b,n),\end{equation}where
\begin{equation}\label{F}
{\cal
L}=2\sqrt{n}\left(\frac{a'b'b}{n}+\frac{ab'^2}{n}+1\right),\end{equation}is
an effective Lagrangian in which primes denotes differentiation
with respect to $r$ and $n$ is given by (\ref{D}). In order to
pass to the Hamiltonian formalism, conjugate momenta must be
calculated. They are given by
\begin{equation}\label{G}
P_a=\frac{\partial {\cal L}}{\partial
a'}=\frac{2bb'}{\sqrt{n}},\hspace{0.5cm}P_b=\frac{\partial {\cal
L}}{\partial b'}=2\frac{(2ab'+a'b)}{\sqrt{n}}.\end{equation}Also,
the primary constraint is given by
\begin{equation}\label{H}
P_n=\frac{\partial {\cal L}}{\partial n'}=0,\end{equation}which is
a consequence of the gauge invariance of general relativity. In
terms of these conjugate momenta the canonical Hamiltonian is
given by
\begin{equation}\label{I}
H=a'P_a+b'P_b-{\cal L},\end{equation}leading to
\begin{equation}\label{J}
H=\sqrt{n}\left(\frac{P_aP_b}{2b}-\frac{a}{2b^2}P_a^2-2\right)=\sqrt{n}{\cal
H}.\end{equation}Because of the existence of constraint (\ref{H}),
the Lagrangian of the system is singular and the total Hamiltonian
can be constructed by adding to $H$ the primary constraints
multiplied by arbitrary functions $\lambda(r)$
\begin{equation}\label{K}
H_T=\sqrt{n}\left(\frac{P_aP_b}{2b}-\frac{a}{2b^2}P_a^2-2\right)+\lambda
P_n.\end{equation}The requirement that the primary constraint
should hold during the $r$-evolution of the system means that
($\approx$ is the Dirac weak equality)
\begin{equation}\label{L}
P'_n=\{P_n,H_T\}\approx 0,\end{equation}where $\{ , \}$ denotes
the Poisson bracket in the minisuperspace $(a,b,n)$ and for any
phase space function can be straightforward obtained by its
standard definition
\begin{equation}\label{L1}
\{f({\bf q},{\bf p}),g({\bf q},{\bf
p})\}=\{\eta^A,\eta^B\}\frac{\partial f}{\partial
\eta^A}\frac{\partial g}{\partial \eta^B},\end{equation}where
$\vec{\eta}=({\bf q};{\bf p})=(a,b,n;P_a,P_b,P_n)$. This leads to
the secondary constraint
\begin{equation}\label{M}
-\frac{1}{2\sqrt{n}}{\cal
H}=-\frac{1}{2\sqrt{n}}\left(\frac{P_aP_b}{2b}-\frac{a}{2b^2}P_a^2-2\right)\approx
0,\end{equation}which represents the invariance of the theory
under $r$-reparametrization. We see that this expression has the
form $f({\bf q},{\bf p})g({\bf q},{\bf p})\approx 0$. As is argued
in \cite{5} and \cite{6} from such relations one cannot conclude
$f({\bf q},{\bf p})\approx 0$ or $g({\bf q},{\bf p})\approx 0$
since this is in contrast with the statement that all secondary
first class constraints must generate symmetries. However, if
$g({\bf q},{\bf p})\approx 0$ then for any function $f({\bf
q},{\bf p})$ we have $f({\bf q},{\bf p})g({\bf q},{\bf p})\approx
0$. As a result, in spite of the usual case where we obtain ${\cal
H}\approx 0$ from the secondary constraint, the constraint
equation (\ref{M}) in this case does not lead us to ${\cal
H}\approx 0$. The Poisson bracket of the secondary constraint with
the total Hamiltonian reads
\begin{equation}\label{N}
\{-\frac{1}{2\sqrt{n}}{\cal H},\sqrt{n}{\cal H}+\lambda
P_n\}=-\frac{\lambda}{2}\{\frac{1}{\sqrt{n}},P_n\}{\cal H}\approx
-\frac{\lambda}{2n}(-\frac{1}{2\sqrt{n}}{\cal H})\approx 0,
\end{equation} which shows that the secondary constraint is
preserved with varying of $r$. Now, following \cite{5}, we define
the extended Hamiltonian as
\begin{equation}\label{O}
H_E=(\sqrt{n}-\frac{\eta}{2\sqrt{n}}){\cal H}+\lambda
P_n,\end{equation}where $\eta$ is is a Lagrange multiplier. The
$r$-evolution of a function is thus given by $f'\approx\{f,H_E\}$.
The preliminary setup for describing the model is now complete. In
what follows, we will study the classical and quantum solutions of
the minisuperspace model described by Hamiltonian (\ref{O}).
\section{Classical solutions}
The classical field equations are governed by the Hamiltonian
equations, that is
\begin{eqnarray}\label{P}
\left\{
\begin{array}{ll}
n'\approx\{n,H_E\}=\lambda,\\\\
P'_n\approx\{P_n,H_E\}=-\frac{1}{2\sqrt{n}}{\cal H}+\frac{\eta}{2n}(-\frac{1}{2\sqrt{n}}{\cal H})=0,\\\\
a'\approx\{a,H_E\}=\frac{1}{\sqrt{n}}(n-\frac{\eta}{2})\left(\frac{1}{2b}P_b-\frac{a}{b^2}P_a\right),\\\\
P'_a\approx\{P_a,H_E\}=\frac{1}{\sqrt{n}}(n-\frac{\eta}{2})\frac{1}{2b^2}P_a^2,\\\\
b'\approx\{b,H_E\}=\frac{1}{\sqrt{n}}(n-\frac{\eta}{2})\frac{1}{2b}P_a,\\\\
P'_b\approx\{P_b,H_E\}=\frac{1}{\sqrt{n}}(n-\frac{\eta}{2})\left(\frac{1}{2b^2}P_aP_b-\frac{a}{b^3}P_a^2\right).
\end{array}
\right.
\end{eqnarray}We see from the first equation of the above system
that the  derivative of $n$ is equal to a Lagrange multiplier.
This shows that $n$ is completely arbitrary and therefore from
(\ref{D}) we have the freedom of choosing $B(r)$ or $N(r)$. Now,
since $n$ can be arbitrarily chosen the rest equations of the
system (\ref{P}) can be cast in the form
\begin{eqnarray}\label{Q}
\left\{
\begin{array}{ll}
a'\approx\{a,H_E\}=\sqrt{n}\left(\frac{1}{2b}P_b-\frac{a}{b^2}P_a\right),\\\\
P'_a\approx\{P_a,H_E\}=\sqrt{n}\frac{1}{2b^2}P_a^2,\\\\
b'\approx\{b,H_E\}=\sqrt{n}\frac{1}{2b}P_a,\\\\
P'_b\approx\{P_b,H_E\}=\sqrt{n}\left(\frac{1}{2b^2}P_aP_b-\frac{a}{b^3}P_a^2\right).
\end{array}
\right.
\end{eqnarray}Up to this point the model, in view of the concerning issue of gauges , has
been rather general and of course under-determined. Before trying
to solve these equations we must decide on a choice of gauge in
the theory. This is important because in order to measure the
physical quantities one should employ gauge conditions. The
under-determinacy problem at the classical level may be removed by
using the gauge freedom via fixing the gauge. From now on we
restrict ourselves to a certain class of gauges, namely $n
=\mbox{const.}$, which is equivalent to the choice $\lambda=0$ in
the first equation of (\ref{P}). With a constant $n$ we assume
$n=1$ without losing general character of the solutions. Although,
it seems that this is the most trivial choice for the function
$\lambda(r)$ in equation (\ref{K}), but this is compatible with
the spirit of gauge invariance, that is, there is no physical
difference between multiple gauge fixing. The measured value of
the system parameters could be different for different gauges but
they have the same behavior. The second and the third equations of
(\ref{P}) can easily be integrated to yield
\begin{equation}\label{R}
b(r)=r,\hspace{0.5cm}P_a=2r,\end{equation}where with the help of
them the rest equations can be put into the form
\begin{eqnarray}\label{S}
\left\{
\begin{array}{ll}
a'=\frac{1}{2r}P_b-\frac{2a}{r},\\\\
P'_b=\frac{P_b}{r}-\frac{4a}{r}.
\end{array}
\right.
\end{eqnarray}Upon integration, the solution of this system can be found as
\begin{equation}\label{T}
a(r)=1-\frac{2M}{r},\hspace{0.5cm}P_b=4-\frac{4M}{r},\end{equation}where
$M$ is an integration constant. Therefore, the above Hamiltonian
formalism lead us to the following static spherically symmetric
metric
\begin{equation}\label{U}
ds^2=-\left(1-\frac{2M}{r}\right)dt^2+N(r)dr^2\pm
2\left[1-\left(1-\frac{2M}{r}\right)N(r)\right]^{1/2}dtdr+r^2d\Omega^2,\end{equation}in
which we have used equation (\ref{D}) to eliminate $B(r)$. This is
exactly the solution obtained in \cite{3} using the Lagrangian
formalism. Now, it is clear that with the choice of
$N(r)=(1-\frac{2M}{r})^{-1}$, we can recover the standard
Schwarzschild black hole solution
\begin{equation}\label{V}
ds^2=-\left(1-\frac{2M}{r}\right)dt^2+\left(1-\frac{2M}{r}\right)^{-1}dr^2+r^2d\Omega^2,\end{equation}with
$M$ being now the mass parameter of the black hole. Other choices
of $N(r)$ are correspond to the other forms of the Schwarzschild
solution, see \cite{3}. It is clear from the condition $a(r)>0$
that this solution is only valid for $r>2M$. The above metric has
an apparent singularity at $r=2M$. This singularity is the
coordinate singularity associated with horizon in the
Schwarzschild space-time, and as is well known, there are other
coordinate system for which this type of singularity is removed
\cite{1}. Another singularity associated with the metric (\ref{V})
is its essential singularity at $r=0$. As we know, in general
relativity, to investigate the types of singularities one has to
study the invariants characteristics of space-time and to find
where these invariants become infinite so that the classical
description of space-time breaks down. In a 4- dimensional
Riemannian space-time there are $14$ independent invariants, but
to detect the singularities it is sufficient to study only three
of them, the Ricci scalar ${\cal R}$, $R_{\mu \nu}R^{\mu \nu}$ and
the so-called Kretschmann scalar $R_{\mu \nu \sigma \delta}R^{\mu
\nu \sigma \delta}$. For the metric (\ref{V}) the Kretschmann
scalar reads
\begin{eqnarray}\label{W}
K=R_{\mu \nu \sigma \delta}R^{\mu \nu \sigma \delta}\sim
\frac{1}{r^6}.
\end{eqnarray}
Now, it is clear that the space-time described by the metric
(\ref{V}) has an essential singularity at $r=0$, which can not be
removed by a coordinate transformation.
\section{Noether symmetry}
As is well known, Noether symmetry approach is a powerful tool in
finding the solution to a given Lagrangian, including the one
presented above. In this approach, one is concerned with finding
the cyclic variables related to conserved quantities and
consequently reducing the dynamics of the system to a manageable
one. The investigation of Noether symmetry in the model presented
above is therefore the goal we shall pursue in this section.
Following \cite{7}, we define the Noether symmetry induced on the
model by a vector field $X$ on the tangent space $TQ=(a, b, a',
b')$ of the configuration space $Q=(a,b)$ of Lagrangian (\ref{F})
through
\begin{equation}\label{AB}
X=\alpha \frac{\partial}{\partial a}+\beta
\frac{\partial}{\partial b}+\frac{d \alpha}{dr}
\frac{\partial}{\partial a'}+\frac{d\beta}{dr}
\frac{\partial}{\partial b'},\end{equation}such that the Lie
derivative of the Lagrangian with respect to this vector field
vanishes
\begin{equation}\label{AC}
L_X {\cal L}=0.\end{equation}In (\ref{AB}), $\alpha$ and $\beta$
are functions of $a$ and $b$ and $\frac{d}{dr}$ represents the Lie
derivative along the dynamical vector field, that is,
\begin{equation}\label{AD}
\frac{d}{dr}=a'\frac{\partial}{\partial
a}+b'\frac{\partial}{\partial b}.\end{equation}It is easy to find
the constants of motion corresponding to such a symmetry. Indeed,
equation (\ref{AC}) can be rewritten as
\begin{equation}\label{AE}
L_X {\cal L}=\left(\alpha \frac{\partial {\cal L}}{\partial
a}+\frac{d\alpha}{dr}\frac{\partial {\cal L}}{\partial
a'}\right)+\left(\beta \frac{\partial {\cal L}}{\partial
b}+\frac{d\beta}{dr}\frac{\partial {\cal L}}{\partial
b'}\right)=0.\end{equation}Noting that $\frac{\partial {\cal
L}}{\partial q}=\frac{dP_q}{dr}$, we have
\begin{equation}\label{AF}
\left(\alpha
\frac{dP_a}{dr}+\frac{d\alpha}{dr}P_a\right)+\left(\beta
\frac{dP_b}{dr}+\frac{d\beta}{dr}P_b\right)=0,\end{equation}which
yields
\begin{equation}\label{AG}
\frac{d}{dr}\left(\alpha P_a+\beta
P_b\right)=0.\end{equation}Therefore, the quantity
\begin{equation}\label{AH}
Q=\alpha P_a+\beta P_b,\end{equation}is constant with respect to
$r$. In terms of the Hamiltonian formalism equation (\ref{AE}) can
be written as
\begin{eqnarray}\label{AI} \alpha \{P_a,{\cal
H}\}+\beta \{P_b,{\cal H}\}+P_a\left[\frac{\partial
\alpha}{\partial a}\{a,{\cal H}\}+\frac{\partial \alpha}{\partial
b}\{b,{\cal H}\}\right]\nonumber \\ +P_b\left[\frac{\partial
\beta}{\partial a}\{a,{\cal H}\}+\frac{\partial \beta}{\partial
b}\{b,{\cal H}\}\right]=0.\end{eqnarray}In order to obtain the
functions $\alpha$ and $\beta$ we use equation (\ref{AI}) (or
(\ref{AE})). In general these equations give a quadratic
polynomial in terms of $P_a$ and $P_b$ (or $a'$ and $b'$) with
coefficients being partial derivatives of $\alpha$ and $\beta$
with respect to the configuration variables $a$ and $b$. Thus, the
resulting expression is identically equal to zero if and only if
these coefficients are zero. This leads to a system of partial
differential equations for $\alpha$ and $\beta$. For Lagrangian
(\ref{F}), condition (\ref{AI}) results in
\begin{eqnarray}\label{AJ}
\alpha\left(\frac{P_a^2}{2b^2}\right)+\beta\left(\frac{P_aP_b}{2b^2}-\frac{a}{b^3}P_a^2\right)+P_a\left[\frac{\partial
\alpha}{\partial
a}\left(\frac{P_b}{2b}-\frac{a}{b^2}P_a\right)+\frac{\partial
\alpha}{\partial b}\frac{P_a}{2b}\right]\nonumber \\
+P_b\left[\frac{\partial \beta}{\partial
a}\left(\frac{P_b}{2b}-\frac{a}{b^2}P_a\right)+\frac{\partial
\beta}{\partial b}\frac{P_a}{2b}\right]=0,\end{eqnarray}which
leads to the following system of equations
\begin{eqnarray}\label{AL}
\left\{
\begin{array}{ll}
\frac{1}{2b^2}\alpha-\frac{a}{b^3}\beta-\frac{a}{b^2}\frac{\partial \alpha}{\partial a}+\frac{1}{2b}\frac{\partial \alpha}{\partial b}=0,\\\\
\frac{1}{2b}\frac{\partial \beta}{\partial a}=0,\\\\
\frac{1}{2b^2}\beta+\frac{1}{2b}\frac{\partial \alpha}{\partial
a}-\frac{a}{b^2}\frac{\partial \beta}{\partial
a}+\frac{1}{2b}\frac{\partial \beta}{\partial b}=0.
\end{array}
\right.
\end{eqnarray}From the second equation of this system we obtain
$\beta=\beta(b)$, that is, the function $\beta$ is independent of
$a$. Then, the third equation gives us
\begin{equation}\label{AM}
\frac{1}{2b^2}\beta+\frac{1}{2b}\frac{\partial \alpha}{\partial
a}+\frac{1}{2b}\frac{d\beta}{db}=0,\end{equation}where upon
integration with respect to $a$ we obtain
\begin{equation}\label{AN}
\alpha(a,b)=-a\left(\frac{1}{b}\beta+\frac{d\beta}{db}\right).\end{equation}Substituting
this result into the first equation of the system (\ref{AL}) we
get $\frac{d^2\beta}{db^2}=0$ and therefore
\begin{equation}\label{AO}
\beta=k_1 b+k_2,\end{equation}where $k_1$ and $k_2$ are
integration constants. Now, it is easy to find $\alpha$ from
(\ref{AN}) as
\begin{equation}\label{AP}
\alpha=-a\left(2k_1+\frac{k_2}{b}\right).\end{equation}The
corresponding constant to the Noether symmetry can be computed
from $Q=\alpha P_a+\beta P_b$. If we use the classical solutions
(\ref{R}) and (\ref{T}) in this relation we obtain $Q=4k_1M+2k_2$.
On the other hand we know that from the point of an observer being
outside the black hole horizon view, its mass is constant and if
one moves along the radial coordinate $r$ this is the only
constant corresponding to the black hole system. Therefore, the
integration constants in (\ref{AO}) can be chosen as
$k_1=\frac{1}{4}$ and $k_2=0$ yielding $\alpha=-\frac{1}{2}a$ and
$\beta=\frac{1}{4}b$ . Hence, the Noether symmetry is generated by
the following vector field
\begin{equation}\label{AQ}
X=-\frac{1}{2}a\frac{\partial}{\partial
a}+\frac{1}{4}b\frac{\partial}{\partial
b}-\frac{1}{2}a'\frac{\partial}{\partial
a'}+\frac{1}{4}b'\frac{\partial}{\partial b'},\end{equation}while,
as a direct consequence of (\ref{AH}), the corresponding Noether
charge is given by
\begin{equation}\label{AS}
Q=-\frac{1}{2}aP_a+\frac{1}{4}bP_b,\end{equation}where its
numerical value is equal to the mass of the black hole $Q=M$.
\section{Quantization of the model}
We now focus attention on the study of the quantization of the
model described above. In the canonical quantization procedure,
the extended Hamiltonian (\ref{O}) contains two first class
constraints (\ref{H}) and (\ref{M}). Therefore, if the wave
function $\Psi(a,b,n)$ describes the quantum version of the
theory, according to the Dirac prescription we demand that it is
annihilated by the operator version of these constraints, that is
\begin{equation}\label{AT}
\hat{P_n}\Psi(a,b,n)=-i\frac{\partial}{\partial
n}\Psi(a,b,n)=0,\end{equation}which implies that the wave function
is independent of $n$. And
\begin{equation}\label{AU}
(\sqrt{n}-\frac{\eta}{2\sqrt{n}})\hat{{\cal
H}}\Psi(a,b)=0\Rightarrow \hat{{\cal
H}}\Psi(a,b)=0,\end{equation}which is known as the Wheeler-DeWitt
equation. Before going any further, some remarks are in order.
Although, the classical equations of motion resulting from the
Lagrangian (\ref{F}) or Hamiltonian (\ref{J}) give the
corresponding classical metric (\ref{U}) or (\ref{V}), the
operator version of the Hamiltonian constraint (\ref{AU}) does not
present a complete description of the quantum version of the
model.  This is because that the Lagrangian (\ref{F}) does not
exhibit the existence of a cyclic variable corresponding to the
Noether symmetry. To be more precise, we seek a point
transformation $(a,b)\rightarrow(u,v)$ on the vector field
(\ref{AB}) such that in terms of the new variables $(u,v)$, the
Lagrangian includes one cyclic variable. A general discussion of
this issue can be found in \cite{8}. Under such point
transformation it is easy to show that the vector field (\ref{AB})
takes the form
\begin{equation}\label{AV}
\tilde{X}=(Xu)\frac{\partial}{\partial
u}+(Xv)\frac{\partial}{\partial
v}+\frac{d}{dr}(Xu)\frac{\partial}{\partial u'}+\frac{d}{dr}(Xv)
\frac{\partial}{\partial v'}.\end{equation}One can show that if
$X$ is a Noether symmetry of the Lagrangian, $\tilde{X}$ has also
this property, that is
\begin{equation}\label{AW}
X{\cal L}=0\Rightarrow \tilde{X}{\cal L}=0.\end{equation}Thus, if
we demand
\begin{equation}\label{AX}
Xu=1,\hspace{.5cm}Xv=0,\end{equation} we get\footnote{The
existence of such a variable which satisfies the condition $Xu=1$
is indeed a consequence of the Noether theorem \cite{Arn}.
However, the change of variables which gives this kind of variable
may be not unique. On the other hand, it is possible that the
system (\ref{AL}) admit more than one set of solutions. In this
case we have more symmetries and therefore more cyclic variables
exist. For instance, if $X_1$ and $X_2$ are two independent
Noether symmetries ($\left[X_1,X_2\right]=0$), we may obtain two
cyclic variables $u^1$ and $u^2$ by solving
$X_1u^1=1,\hspace{0.2cm} X_1u^{i\neq 1}=0$ and $X_2
u^2=1,\hspace{0.2cm} X_2u^{i\neq 2}=0$. }
\begin{equation}\label{AY}
\tilde{X}=\frac{\partial}{\partial u}\Rightarrow\tilde{X}{\cal
L}=\frac{\partial {\cal L}}{\partial u}=0.\end{equation}This means
that $u$ is a cyclic variable and the dynamics can be reduced. On
the other hand, one can show that \cite{9}, the constant of motion
$Q$ which corresponds to the Noether symmetry is nothing but the
momentum conjugated to the cyclic variable, that is, $Q=P_u$.  To
find the explicit form of the above mentioned point transformation
we should solve the equations (\ref{AX}), which give
\begin{equation}\label{AZ}
-\frac{1}{2}a\frac{\partial u}{\partial
a}+\frac{1}{4}b\frac{\partial u}{\partial b}=1,\end{equation}
\begin{equation}\label{BA}
-\frac{1}{2}a\frac{\partial v}{\partial
a}+\frac{1}{4}b\frac{\partial v}{\partial b}=0.\end{equation}
These differential equations admit the following general solutions
\begin{equation}\label{BC}
u(a,b)=\ln \frac{b^2}{a}
+f_1(ab^2),\hspace{0.5cm}v(a,b)=f_2(ab^2),\end{equation}where
$f_1$ and $f_2$ are two arbitrary functions of $ab^2$. As is
indicated in \cite{8}, "the change of coordinates is not unique
and a clever choice is always important". With a glance at the
Lagrangian (\ref{F}), we choose the functions $f_1$ and $f_2$ as
\begin{equation}\label{BD}
f_1(ab^2)=f_2(ab^2)=\ln (ab^2).\end{equation}With this choice, the
Lagrangian (\ref{F}) takes the form
\begin{equation}\label{BE}
{\cal
L}=e^v\left(-\frac{1}{8}u'^2+\frac{1}{2}u'v'\right)+2.\end{equation}It
is clear from this Lagrangian that $u$ is cyclic and the Noether
symmetry is given by $P_u=Q=\mbox{const}$. Also, the momenta
conjugate to $u$ and $v$ are
\begin{equation}\label{BF}
P_u=\frac{\partial {\cal L}}{\partial
u'}=e^v\left(-\frac{1}{4}u'+\frac{1}{2}v'\right),\hspace{0.5cm}P_v=\frac{\partial
{\cal L}}{\partial v'}=\frac{1}{2}u'e^v,\end{equation}which give
rise to the following Hamiltonian for our model
\begin{equation}\label{BG}
{\cal
H}=e^{-v}\left(2P_uP_v+\frac{1}{2}P_v^2\right)-2.\end{equation}Since
this Hamiltonian and $P_u$ commute with each other  $[{\cal
H},P_u]=0$, they have simultaneous eigenfunctions. Therefore, the
quantum description of our Noether symmetric black hole model can
be viewed by the following equations
\begin{equation}\label{BH}
{\cal
H}\Psi(u,v)=\left[e^{-\hat{v}}\left(2\hat{P_u}\hat{P_v}+\frac{1}{2}\hat{P_v}^2\right)-2\right]\Psi(u,v)=0,\end{equation}
\begin{equation}\label{BI}
\hat{P_u}\Psi(u,v)=Q\Psi(u,v).\end{equation}Choice of the ordering
$e^{-\hat{v}}\hat{P_v}\rightarrow
\frac{1}{2}(e^{-\hat{v}}\hat{P_v}+\hat{P_v}e^{-\hat{v}})$ and
$e^{-\hat{v}}\hat{P_v}^2\rightarrow
\hat{P_v}e^{-\hat{v}}\hat{P_v}$ to make the Hamiltonian Hermitian
and use of $\hat{P_u}\rightarrow -i \partial_u$ and similarly for
$\hat{P_v}$ the above equations read
\begin{equation}\label{BJ}
\left(-\frac{1}{2}\frac{\partial ^2}{\partial
v^2}+\frac{1}{2}\frac{\partial}{\partial
v}-2\frac{\partial^2}{\partial u \partial
v}+\frac{\partial}{\partial
u}-2e^v\right)\Psi(u,v)=0,\end{equation}
\begin{equation}\label{BK}
-i\frac{\partial}{\partial
u}\Psi(u,v)=Q\Psi(u,v).\end{equation}The solutions of the above
differential equations are separable and may be written in the
form $\Psi(u,v)= U(u)V(v)$. Equation (\ref{BK}) can be immediately
integrated leading to a oscillatory behavior for the wave function
in $u$ direction, i.e. in the direction of symmetry, that is
\begin{equation}\label{BL}
\Psi(u,v)=e^{iQu}V(v).\end{equation}Substitution of this result
into relation (\ref{BJ}) yields the following equation for the
function $V(v)$
\begin{equation}\label{BM}
\left[\frac{d^2}{dv^2}+(4iQ-1)\frac{d}{dv}+(4e^v-2iQ)\right]V(v)=0,\end{equation}where
its solutions can be written in terms of the Hankel (or Bessel)
functions as
\begin{equation}\label{BN}
V(v)=e^{(\frac{1}{2}-2iQ)v}\left[c_1H^{(1)}_{i\sqrt{16Q^2-1}}\left(4e^{v/2}\right)+c_2H^{(2)}_{i\sqrt{16Q^2-1}}\left(4e^{v/2}\right)\right].\end{equation}Thus,
the eigenfunctions of the Wheeler-DeWitt equation can be written
as
\begin{equation}\label{BO}
\Psi_Q(u,v)=e^{v/2}e^{iQ(u-2v)}\left[c_1H^{(1)}_{i\sqrt{16Q^2-1}}\left(4e^{v/2}\right)+c_2H^{(2)}_{i\sqrt{16Q^2-1}}\left(4e^{v/2}\right)\right].\end{equation}We
may write the general solutions to the Wheeler-DeWitt equation as
a suitable superposition of these eigenfunctions. In the classical
limit, i.e. for large values of $r$ we have $b(r)\sim r$ and
$a(r)\sim 1$ and from equations (\ref{BC}) and (\ref{BD}) we get
the behavior $u(r)\sim \ln r^4$ and $v(r)\sim \ln r^2$ for $u$ and
$v$ in this limit. On the other hand, in view of the
asymptotically behavior of the Hankel functions
$H^{(1)(2)}_{\nu}(z) \sim z^{-1/2}e^{\pm i[z-(2\nu+1)\pi /4]}$, we
obtain the following behavior for the Wheeler-DeWitt eigenfunction
for large values of $r$
\begin{equation}\label{BP}
\Psi_Q(u,v) \sim e^{v/4}e^{\pm
i(Qu-2Qv+4e^{v/2})},\end{equation}Therefore, in the semiclassical
approximation region we obtain the phase function $S(u,v)$ as
\begin{equation}\label{BQ}
S(u,v)=\pm \left(Qu-2Qv+4e^{v/2}\right).\end{equation}where the
positive sign corresponds to an increasing $r$ model. In the WKB
method, the correlation between classical and quantum solutions is
given by the relation $P_q=\frac{\partial S}{\partial q}$ . Thus,
using the definition of $P_u$ and $P_v$ in (\ref{BF}), the
equation for the classical trajectories becomes
\begin{eqnarray}\label{BR}
\left\{
\begin{array}{ll}
e^{v}\left(-\frac{1}{4}u'+\frac{1}{2}v'\right)=Q,\\\\
\frac{1}{2}u'e^v=-2Q+2e^{v/2}.
\end{array}
\right.
\end{eqnarray}Eliminating $u'$ from this system results
$v'e^{v/2}=2$, which can easily be integrated to yield $v(r)=\ln
r^2$. With the help of this expression for $v$, the equation for
$u'$ can be put into the form $u'=-\frac{4Q}{r^2}+\frac{4}{r}$,
which admits the solution $u(r)\sim \ln r^4$ for large values of
$r$. Therefore, our analysis shows that in the large-$r$ limit the
behavior of the classical solution is exactly recovered. The
meaning of this result is that for large values of $r$ the
effective action is very large and the system can be described
classically. On the other hand, near the black hole singularities
we cannot neglect the quantum effects and the classical
description breaks down. Since the WKB approximation is no longer
valid in this regime, one should go beyond the semiclassical
approximation. In the cases where $r$ approaches the classical
singularities $r\rightarrow 0$ and $r\rightarrow 2M$ the variables
$u$ and $v$ behave as $u,v\rightarrow -\infty$ and $u\rightarrow
\ln (2M)^4$, $v\rightarrow -\infty$ respectively. We see from
(\ref{BO}) that the Hankel functions have small argument in these
limits with a oscillatory behavior. Therefore, since the
eigenfunctions tend to zero, the quantum solutions are regular for
$r\rightarrow 0, 2M$. These quantum solutions may be interpreted
as being responsible of the avoidance of classical singularity. We
are not here representing a complete discussion about this subject
since it depends on the various aspects of the Wheeler-DeWitt
equation such as the choice of ordering, representation for the
momenta and choice of suitable superposition of its eigenfunctions
to construct the wave packets.
\section{Conclusions}
In this paper we have studied a static spherically symmetric
space-time in a canonical point of view. To construct the
Hamiltonian formalism we have followed a method proposed in
\cite{3} in which a $3+1$ decomposition is performed with respect
to the radial coordinate $r$. The resulting $r$-Hamiltonian is
shown to satisfy the Hamiltonian constraint which represents that
the underlying theory is invariant under $r$-reparametrization. We
have shown that the classical Hamiltonian equations admit
solutions which can be identified with the Schwarzschild black
hole. The existence of Noether symmetry implies that the Lie
derivative of the Lagrangian with respect to the infinitesimal
generator of the desired symmetry vanishes. By applying this
condition to the Lagrangian of the model, we showed that the mass
parameter of the black hole plays the role of the Noether charge,
i.e. the mass is a constant of $r$-motion. We have then quantized
the model and shown that the corresponding quantum black hole
model and the ensuing Wheeler-DeWitt equation are amenable to
exact solutions in terms of the Hankel functions. In classical
regime we have seen that these solutions are expressed in terms of
a superposition of states of the form $e^{iS}$ due to the
existence of Noether symmetry. In semiclassical approximation for
quantum gravity, this type of state represents the correlations
between classical trajectories and the quantum wave function.
Using this interpretation we have shown that the corresponding
classical metric can be recovered by the quantum solutions
counterparts. Finally, we have presented a short discussion to
clear the regular behavior of the wave function near the classical
singularities and the possibility of the avoidance of these
singularities due to quantum effects.

\end{document}